\documentclass[10pt]{pazh}
\usepackage{graphicx}

\def\farcs{\hbox{$.\!\!^{\prime\prime}$}}
\def\widerul{\vrule height 2.5ex width 0ex depth 1.0ex}
\def\text{\mbox}
\def\etal{{\it et\,al.}}
\begin{document}

\title{NEW STUDIES OF THE PULSAR WIND NEBULA \protect\\ IN THE SUPERNOVA
REMNANT CTB\,80}

\author{T.~A.~Lozinskaya\inst{1}$^{^*}$,
V.~N.~Komarova\inst{2},
A.~V.~Moiseev\inst{2},
S.~I.~Blinnikov\inst{2}}
\date{Received --- November~15, 2004}

\titlerunning{NEW STUDIES OF THE PULSAR WIND NEBULA}

\authorrunning{LOZINSKAYA et al.}

\institute{Sternberg Astronomical Institute,
Universitetskii pr.~13, Moscow, 119992 Russia
\and Special Astrophysical Observatory, Russian Academy of Sciences, Nizhnii
Arkhyz, Karachai-Cherkessian Republic, 357147 Russia
\and Institute for Theoretical and Experimental Physics, ul. Bol'shaya
Cheremushkinskaya 25, Moscow, 117259 Russia}
\email{lozinsk@sai.msu.ru}

\abstract
{We investigated the kinematics of the pulsar wind nebula (PWN) associated with
PSR~B1951$+$32 in the old supernova remnant CTB\,80 using the Fabry--Perot
interferometer of the 6\,m Special Astrophysical Observatory telescope. In
addition to the previously known expansion of the system of bright filaments
with a velocity of 100--200~km~s$^{-1}$, we detected weak high-velocity
features in the H$\alpha$ line at least up to velocities of 400--450~km~s$^{-1}$.
We analyzed the morphology of the PWN in the H$\alpha$, [SII], and [OIII]
lines using HST archival data and discuss its nature. The shape of the central
filamentary shell, which is determined by the emission in the [OIII] line and
in the radio continuum, is shown to be consistent with the bow-shock model for
a significant (about~$60^\circ$) inclination  of the pulsar's velocity vector
to the plane of the sky. In this case, the space velocity of the pulsar is
twice higher than its tangential velocity, i.e., it reaches $\simeq 500$~km~s$^{-1}$,
and PSR~B1951$+$32 is the first pulsar whose line-of-sight velocity (of about
400 km/s) has been estimated from the PWN observations. The shell-like
H$\alpha$-structures outside the bow shock front in the east and the west may
be associated with both the pulsar's jets and the pulsar-wind breakthrough
due to the layered structure of the extended CTB\,80 shell.
}

\maketitle

\section*{INTRODUCTION}

CTB\,80 is a classical example of a supernova remnant (SNR) with a fast-moving
pulsar at a late stage of the pulsar's interaction with a very old shell.

The radio~image of CTB\,80 is represented by three ex-\\tended (about 30$'$)
ridges that converge in the region of a bright compact core (Velusami and
Kundu\,1974; Velusami \etal\,1976; Angerhofer \etal\,1981; Strom \etal\,1984;
Mantovani \etal\,1985; Strom~1987; Castelleti \etal\,2003, and references
therein). This unusual (for SNRs) morphology ceased to appear puzzling in
1988, when, on the one hand, the 39.5-ms pulsar PSR~B1951$+$32 was discovered
in the core (Kulkarni \etal\,1988; Fruchter \etal\,1988) and, on the other
hand, Fesen \etal\,(1988) identified an extended infrared shell that extends
the radio~ridges in the northeast and assumed that precisely this infrared
shell represents a very old SNR. The radio~ridges with a steep spectrum
($\alpha \simeq -0.7$) correspond to the part of the shell into the
compressed magnetic field of which the approached pulsar injected fresh
relativistic particles, thereby reanimating its synchrotron radio emission.
Subsequently, an H\,I shell that coincides with the infrared shell and expands
with a velocity of 72~km~s$^{-1}$, which yields a kinematic SNR age of
$7.7\times 10^{4}$\,yr, was also identified (Koo \etal\,1990, 1993).
The characteristic age of the pulsar PSR~B1951$+$32 is $t\simeq 10^5$\,yr
(Kulkarni \etal\,1988; Fruchter \etal\,1988).

The pulsar wind nebula (PWN) is represented by a bright compact ($45''$) core
with a flat spectrum ($\alpha \simeq 0.0$). The core lies at the western
boundary of a $10' \times 6'$ plateau with a steeper spectrum ($\alpha \simeq
-0.3$) elongated in the east--west direction. Recently, Migliazzo \etal\,(2002)
detected the motion of the pulsar with a velocity of 240~km~s$^{-1}$ in a
direction that confirms the possibility of its birth inside the infrared and
H\,I shells.

The stage of pulsar wind interaction with a very old shell observed in CTB\,80
seems most complex in the evolutionary chain of PWNs, from young SNRs where
the pulsar interacts with the SN ejection to pulsars that have already
escaped from the old SNR. The difficulty lies in the fact that the matter
density in the cooled old shell behind the front of the decelerated shock
(and, hence, the density of the compressed interstellar magnetic field) is
several hundred times higher than the ambient density and the shell structure
is unpredictable in advance. In addition, the rotational energy loss by the
pulsar PSR~B1951$+$32 is large ($\dot E =3.7\times 10^{36}$~erg~s$^{-1}$, as
estimated by Kulkarni \etal\,1988), which is comparable to the energy input
from the young Vela~X pulsar.

Various distance estimates for CTB\,80 and the pulsar PSR~B1951$+$32 lie
within the range 1.5\mbox{--}2.5~kpc (see Koo \etal\,1993; Strom and
Stappers\,2000; and references therein); we use the universally accepted
distance of 2~kpc.

In this paper, we present our observations of the PWN in CTB\,80 with
the 6\,m Special Astrophysical Observatory (SAO) telescope and analyze
narrow-band observational data from the HST archive.

\section*{OBSERVATIONS AND DATA REDUCTION}

\subsection*{Interferometric Observations on the 6\,m Telescope}

The core of CTB\,80 was observed with the 6\,m telescope as part of a program
entitled ``The Kinematics of Matter in Pulsar Wind Nebulae'' (the main
applicant is Yu.A.~Shibanov). Our interferometric observations of CTB\,80 were
performed on October 9--10, 2001, at the prime focus of the 6\,m telescope
using the SCORPIO focal reducer; the equivalent focal ratio of the system was
$F/2.9$. A description of SCORPIO is given in the paper by Afanasiev and
Moiseev~(2005) and on the Internet
(http://www.sao.ru/hq/moisav/scorpio/scor\-pio.html); the SCOPRIO capabilities
in interferometric observations were described by Moiseev~(2002). The seeing
during the
observations varied within the range $1\farcs 2\text{--}2\farcs 2$. We
used a scanning Fabry--Perot interferometer (FPI) in the 235th order at the
wavelength of the H$\alpha$ line; the separation between the neighboring orders
of interference, $\Delta\lambda=28$~\AA, corresponded to a region free from
order overlapping of 1270~km~s$^{-1}$ on the radial velocity scale. The width
of the instrumental FPI profile
was $FWHM\approx2.5$~\AA\ or $\simeq 110\text{--}120$~km~s$^{-1}$.
Premonochromatization was performed using an interference filter with
a half-width of $\Delta\lambda =14$~\AA\ centered on the H$\alpha$ line.
The detector was a TK1024 $1024\times1024$-pixel CCD~array. The observations
were carried out with $2\times2$-pixel binning to reduce the readout time. In
each spectral channel, we obtained $512\times512$-pixel images; at a
$0.55''$/\mbox{pixel} scale, the total field of view was $4\farcm7$. We
obtained
a total of 32 interferograms at various FPI plate spacings, so the width of
the spectral channel was $\delta \lambda=0.87$~\AA\ or 40~km~s$^{-1}$ near
H$\alpha$. The exposure time was 240~s per channel.

We reduced the observations using software running in the IDL environment
(Moiseev\,2002). After the primary reduction (debiasing and flat fielding),
the observational data were represented as $512\times512\times32$-pixel data
cubes; here, a 32-channel spectrum corresponds to each pixel. Major
difficulties in the data reduction process arose when night-sky emission lines
were subtracted from interferograms. All bright image features were masked,
and an azimuthally averaged radial sky emission line profile was constructed
from the remaining areas; this profile was subtracted from the corresponding
frame in the data cube. This technique allows the sky line intensity
variations during two-hour observations to be effectively corrected. However,
almost the entire field of view near the core of CTB\,80 proved to be filled
with weak emission features, which was also confirmed by our deep H$\alpha$
images of the PWN. Therefore, the problem of choosing a sufficient number of
``clean'' sky areas in a region of almost~$5'$ in size arose. As a result,
the weakest features in the field of view could be ``oversubtracted'', and
the H$\alpha$ emission line profile is severely distorted at these locations.
Our estimates indicate that the line distortion due to the background
subtraction may be disregarded for regions in which the intensity of the
H$\alpha$ line exceeds 150~photoelectrons per CCD~pixel, which corresponds
to a surface brightness of
$8.6\times 10^{-17}$~erg~s$^{-1}/\sq^{''}$ (uncorrected for the interstellar
reddening).

\begin{table*}[t!]
\centerline{Details of the narrow- and medium-band observations of the CTB\,80 core with the
6\,m telescope (BTA) and HST}
\vspace*{0.3cm}
\hspace*{2.5cm}
{\large
\begin{tabular}{l|l|c|c|c}
\hline \multicolumn{1}{c|}{Range, telescope}& \multicolumn{1}{c|}{Filter}&$\lambda_{\textrm{cen}}$,
\AA &
 \multicolumn{1}{c|}{$FWHM$, \AA} &Exposure time, s\widerul\\
\hline
H$\alpha$, BTA& FN657 &   $6578$    &  75   & $4\times300$  \widerul \\
Continuum, BTA& SED707&   $7036$    &  207  &  $4\times120$   \widerul\\
\hline
H$\alpha$, HST & F656N& $6563$ & 20  & $2600+2700$\widerul\\
Continuum, HST & F547M& $5479$ & 200 & $1300+1300$\widerul\\
$[$OIII$]$, HST & F502N& $5013$ & 20  & $2700+2700$\widerul\\
$[$SII$]$, HST & F673N& $6732$ & 20  & $2700+2700$\widerul \\
\hline
\end{tabular}
}
\end{table*}

The large number of background stars in the field allowed us to measure and
correct the atmospheric transparency and seeing variations in each image.
The resulting seeing was
$2\farcs2$. Subsequently, we reduced the spectra in the data cube to the same
wavelength scale. The formal accuracy of measuring the relative line-of-sight
velocities was about 2--4~km~s$^{-1}$. However, since the emission lines in
the object often have asymmetric profiles, the actual measurement accuracy
depends on the chosen line profile fitting method.

The spectra of the object were smoothed with a $FWHM=1.7$~\AA\ (two channels)
Gaussian for optimal filtering of the data cube. To reliably identify weak
emission features, we also smoothed the images in the cube with a bivariate
Gaussian. The resultant angular
resolution was about $2\farcs6$. The smoothing was performed using the ADHOC
software package\footnote{The ADHOC software package was developed by
J.~Boulestex (Marseilles Observatory) and is freely accessible on the
Internet.}.

The continuum level in each spectrum was determined as a median mean of
the eight weakest levels. When constructing the velocity field and the
monochromatic image in the H$\alpha$ line, we fitted the spectral line
profile by a Gaussian using only the points offset by $\pm3$ channels from
the predetermined line peak.

We calibrated the monochromatic image in energy units (erg~s$^{-1}$~cm$^{-2}$)
by comparison with our calibrated narrow-band image (see below).

All of the radial velocities in this paper are heliocentric; the passage to
the Local Standard of Rest corresponds to $V_{\textrm{Hel}}=V_{\textrm{LSR}}
+ 17.6$~km~s$^{-1}$.

\subsection*{Medium- and Narrow-Band Images}

\textbf{Observations with the 6\,m telescope.} Our medium- and narrow-band
observations of the CTB\,80 core were performed on October~11 and~12, 2001,
with the 6\,m telescope using the SCORPIO focal reducer (see above) at seeing
$1\farcs4\text{--}1\farcs6$ at zenith distance
$z=12^{\circ}$--$21^{\circ}$. Filter parameters and exposure times are given
in the Table.

To calibrate the images, we observed the spectrophotometric standards G138--31
and Feige~110.

\textbf{The HST archive.} The PWN in CTB\,80 was observed with HST (the main
applicant is J.~Trauger) in October~1997 using the WFPC2 instrument.
Parameters of the filters used and the total exposure time are given in the
Table. The primary data reduction is automatically performed when the
data are queried from the HST archive. A filtering code proposed by
N.A.~Tikhonov (private communication) was used to remove numerous cosmic-ray
particle hits. The formal accuracy of the astrometric referencing using
WCSTools is an order of magnitude lower than the internal accuracy of
the USNO-A2.0 star catalog used
as a reference one. The resultant value was taken to be $0\farcs3$.

\section*{RESULTS OF OBSERVATIONS}

\subsection*{The PWN Morphology}

Previous studies of the CTB\,80 core showed that its optical emission is
typical of SNRs: an intense (relative to H$\alpha$) [NII], [SII], [OI], [OIII]
line emission and a filamentary structure (see Angerhofer \etal\,1980; Blair
\etal\,1984; Whitehead \etal\,1989; Hester and Kulkarni~1988, 1989; and
references therein). Significant differences in the PWN morphology in
different lines were reported: only a symmetric central shell is observed in
the [OIII] line, a shell to the east of it appears in the [SII] line, and two
bright shells to the east and the west of the central structure that form an
elongated core structure are observed in the H$\alpha$ line.

The deep HST images of the PWN demonstrate a staggering  filamentary,
irregularly shaped multishell structure and confirm the results of
ground-based observations (see Fig.\,1, which shows the [OIII], [SII],
and H$\alpha$ images, from top to bottom panels, respectively, with
superimposed VLA 1.5-GHz radio isophotes).

Only the central horseshoe-shaped part of the core closest to the pulsar is
seen in the [OIII] line. The PWN morphology in the [OIII] line clearly shows
a structure expected for the bow shock produced by the pulsar's motion (see
the discussion). The H$\alpha$ emission is also observed in this central
filamentary horseshoe-shaped shell; the H$\alpha$ filaments are adjacent to
the [OIII] filaments from the outside. We clearly see the two bright (in
H$\alpha$) filamentary shells in the east and the west, which form a
spindle-shaped core structure in the east--west direction. The PWN sizes in
the H$\alpha$ line are about $75''\times 38''$ or $0.73\times0.37$~pc.

\subsection*{The PWN Kinematics}

\textbf{The velocities of PWN filaments.} The comparisons of relative line
intensities in the core spectrum with diagnostic models made by several
authors are suggestive of collisional excitation of the gas behind the front
of a shock propagating at a velocity of about 120\mbox{--}140~km~s$^{-1}$ in
a medium with a density of 25--100~cm$^{-3}$ (Hester and Kulkarni~1989, and
references therein). Gas motions in the PWN with such velocities have already
been detected. The main method used was long-slit spectroscopy, which allowed
the velocities of the individual, generally brightest (and, hence,
possibly slowest) knots and filaments to be estimated (Angerhofer \etal\,1981;
Blair \etal\,1984). Spectroscopic methods did not reveal radial velocities
higher than 200~km~s$^{-1}$ with a surface brightness
at H$\alpha \geq 0.5\ 10^{-6}$~erg~cm$^{-2}$~s$^{-1}$~sr$^{-1}$
anywhere in the PWN (Blair \etal\,1988). Using the echelle spectrograph of
the 4-m WHT telescope, Whitehead \etal\,(1989) found an expansion of the two
central shells with velocities of~200 and 100~km~s$^{-1}$.

Using FPI, we have studied the PWN kinematics for the first time. The
advantage of FPI observations is that they give the line profile everywhere
in the PWN, and not only in the region cut out by the spectrograph slit.

Our observations indicate that the line profile, which is single in bright
peripheral filaments, is characterized by a multicomponent structure or line
asymmetry, bright red wings in the central PWN regions. As an illustration,
several profiles and their decomposition into individual Gaussians are given
in Fig.\,2. Having analyzed all of the observed H$\alpha$ profiles, we were
able to localize the high-velocity gas in the PWN image that emits in the
range from~200 to 400~km~s$^{-1}$ (shown at the center in Fig.\,2).

\begin{figure}[t!]
\vspace*{0.05cm}
\includegraphics[width=8.5cm,bb=173 67 417 676,clip]{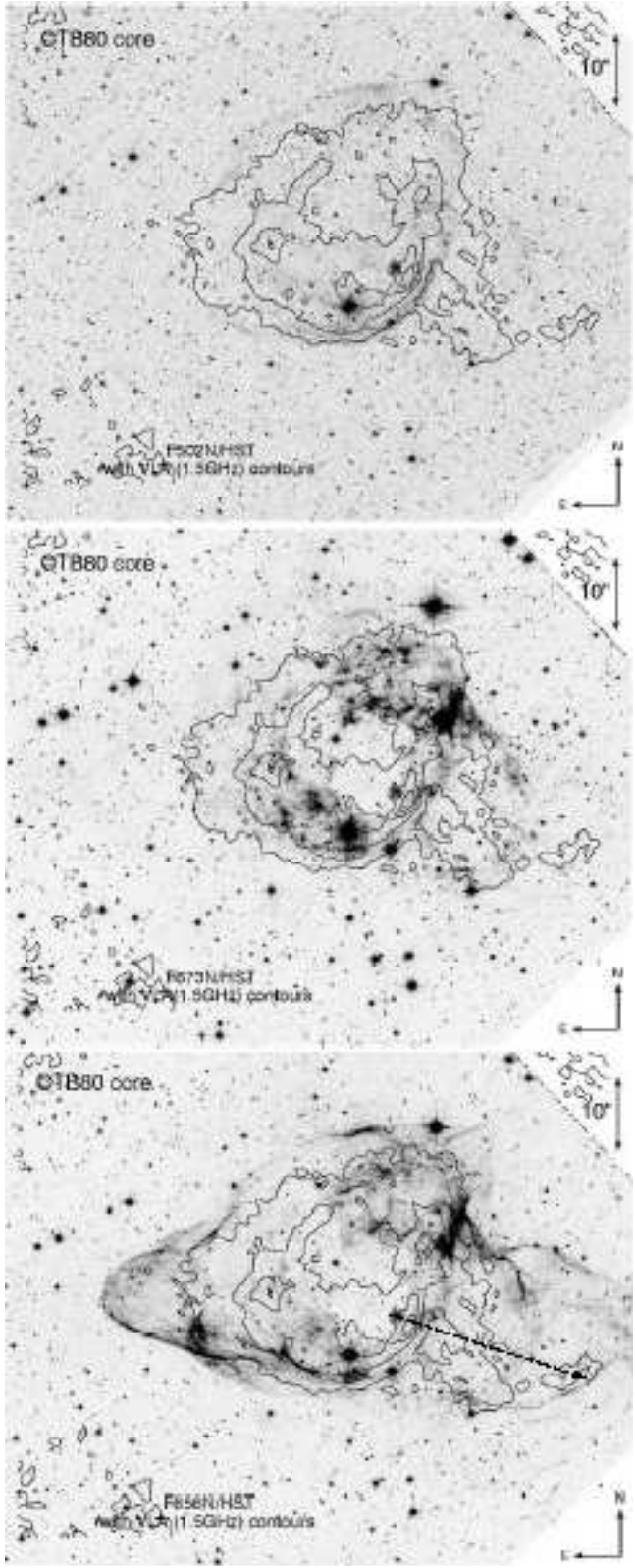}
\caption{HST images of the PWN with superimposed VLA 1.5-GHz radio isophotes:
in the [OIII] line ({\it top}), in the [SII] line ({\it middle}), and in
the H$\alpha$ line ({\it bottom panel}). The arrow indicates the motion of
the pulsar over a period of 1000\,yr, as measured by Migliazzo \etal\,(2002).
\hfill}
\end{figure}

Figure~3 shows several so-called position--velocity diagrams (the change in
gas velocity along the chosen direction) constructed from our FPI observations
and the localization of the corresponding directions in the PWN image.

As we see from this figure, in the bright core structures,
the position--velocity diagrams show velocities that agree with
the previously measured expansion velocity of 100--200~km~s$^{-1}$; i.e.,
they completely confirm the results of spectroscopic observations.

In addition to the gas velocities in bright filaments within the range
from~$-200$ to~$+200$~km~s$^{-1}$, our observations clearly reveal weaker
high-velocity H$\alpha$ emission features in the PWN. Weak emission at high
velocities is clearly seen in the position--velocity diagrams up to
400--450~km~s$^{-1}$. This is a lower limit, since the velocity range is
limited by the FPI velocity range free from order overlapping and the
impossibility to properly take into account the contribution from the [NII]
6548\,\AA\ line emission.

Figure~3 (scan~3) clearly shows the typical (for an expanding shell)
structure of the so-called velocity ellipsoid (its half corresponding to
positive velocities) that is determined by high-velocity motions. The
corresponding possible expansion velocity reaches the lower limit mentioned
above.

The surface brightness of the detected high-velocity component of the
H$\alpha$ line in the central part of the core lies within the range
$(1.4\text{--} 20)\times 10^{-16}$~erg~s$^{-1}$~cm$^{-2}$~arcsec$^{-2}$. In
our estimation, we took the color excess $E(B-V)=0.8$ from Blair \etal\,(1984).

Figure~4 shows the line-of-sight velocity field corresponding to the peak of
the main line component constructed from our observations and superimposed on
the H$\alpha$ image of the PWN. We emphasize that the line-of-sight velocity
field refers only to the core of the line, which often has a complex
multicomponent profile. The velocities determined from the line peak lie
within the range $-100$ to $+50$~km~s$^{-1}$, in agreement with the
spectroscopic observations.

In Fig.\,4, we clearly see the symmetry axis in the velocity
distribution of the peak of the main line component in a direction
$P\approx 230^\circ$, which is close to but does not coincide with
the direction of the pulsar's motion ($P = 252^\circ \pm 7^\circ$;
Migliazzo \etal\,2002).

We estimated the total flux from the PWN in the main line component to be
$8.4\times 10^{-12}$~erg~s$^{-1}$~cm$^{-2}$, in close agreement with the
estimate of Whitehead \etal\,(1989). The total flux in the high-velocity line
component is $6.4\times 10^{-13}$~erg~s$^{-1}$~cm$^{-2}$; the luminosity of
the high-velocity gas accounts for about~7\% of the total luminosity of
the core in the H$\alpha$ line.

The high-velocity emission detected in the core for the first time has
confirmed the changes in the PWN fine structure pointed out by Strom and
Blair\,(1985) --- possibly the proper motions of the filaments of the central
shell corresponding to a velocity of $250 d$~km~s$^{-1}$ (up to
$400 d$~km~s$^{-1}$), where~$d$ is the distance to CTB\,80 in
kpc, i.e., about 500~km~s$^{-1}$ (possibly up to 800~km~s$^{-1}$).

\begin{figure*}[t!]
\vspace*{0.05cm}
\hspace*{-0.25cm}
\includegraphics[width=18.25cm,bb=30 90 575 585,clip]{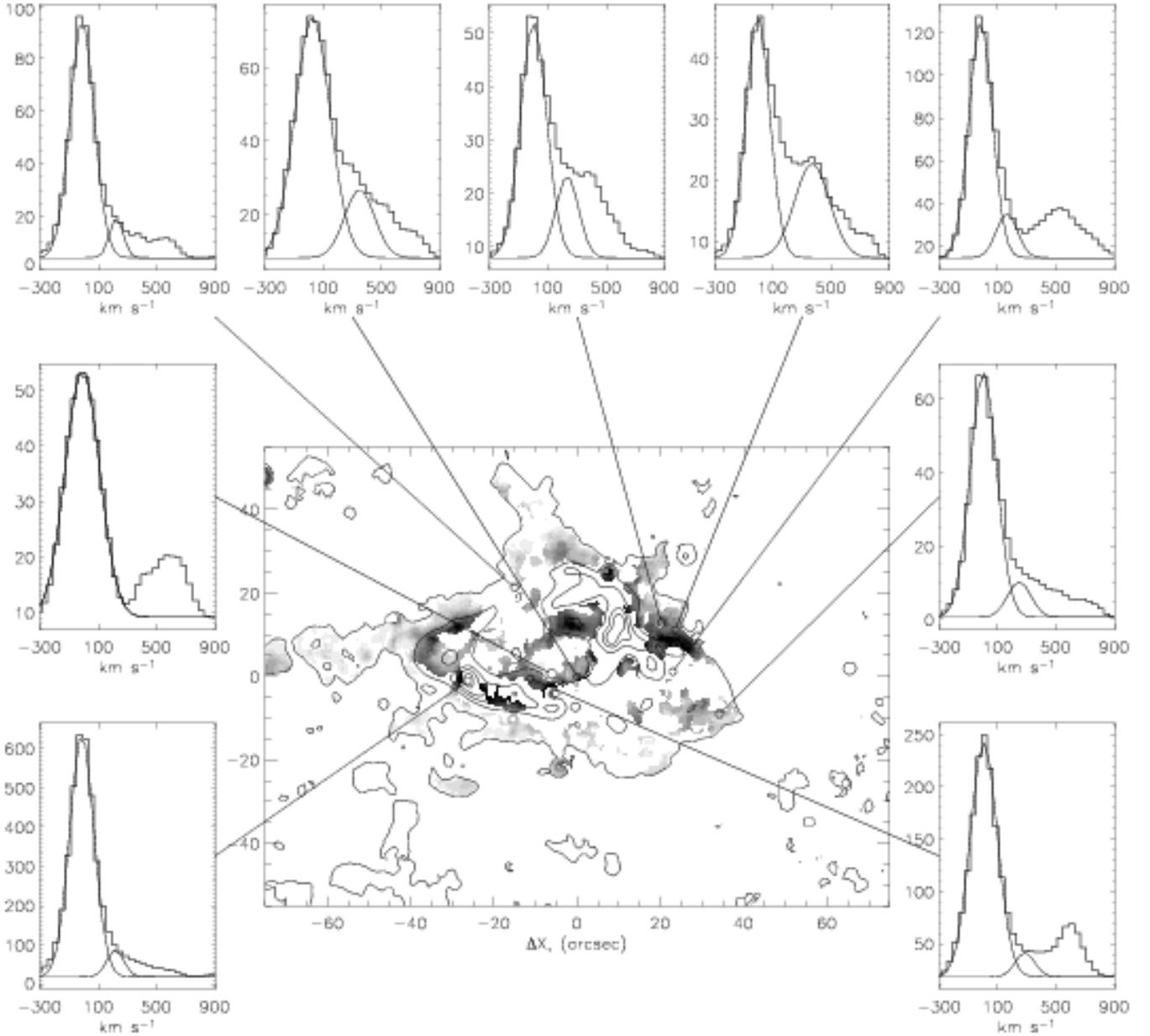}
\caption{H$\alpha$ profiles at several points of the PWN and their Gaussian
decomposition. The features at a velocity of $+605$~km~s$^{-1}$ are
attributable to the [NII] 6548\,\AA\ line emission that falls in the wings of
the interference filter and that is separated from the H$\alpha$ emission by
about 0.5 order of interference. The localization of the high-velocity gas
emitting in the range from~200 to 400~km~s$^{-1}$
is shown in the central H$\alpha$ image of the PWN (isophotes). \hfill}
\end{figure*}

\textbf{The velocities of filaments in the CTB\,80 shell outside the core.}
The optical radiation from the extended SNR CTB\,80 outside the PWN is
represented by the system of faint filaments clearly seen in the [SII] and
H$\alpha$ lines over the entire $16'\times16'$ field whose image is given in
the paper by Hester and Kulkarni~(1989).

\begin{figure*}[t!]
\vspace*{0.05cm}
\hspace*{2.0cm}
\includegraphics[height=20.15cm,bb=30 -20 485 740,clip]{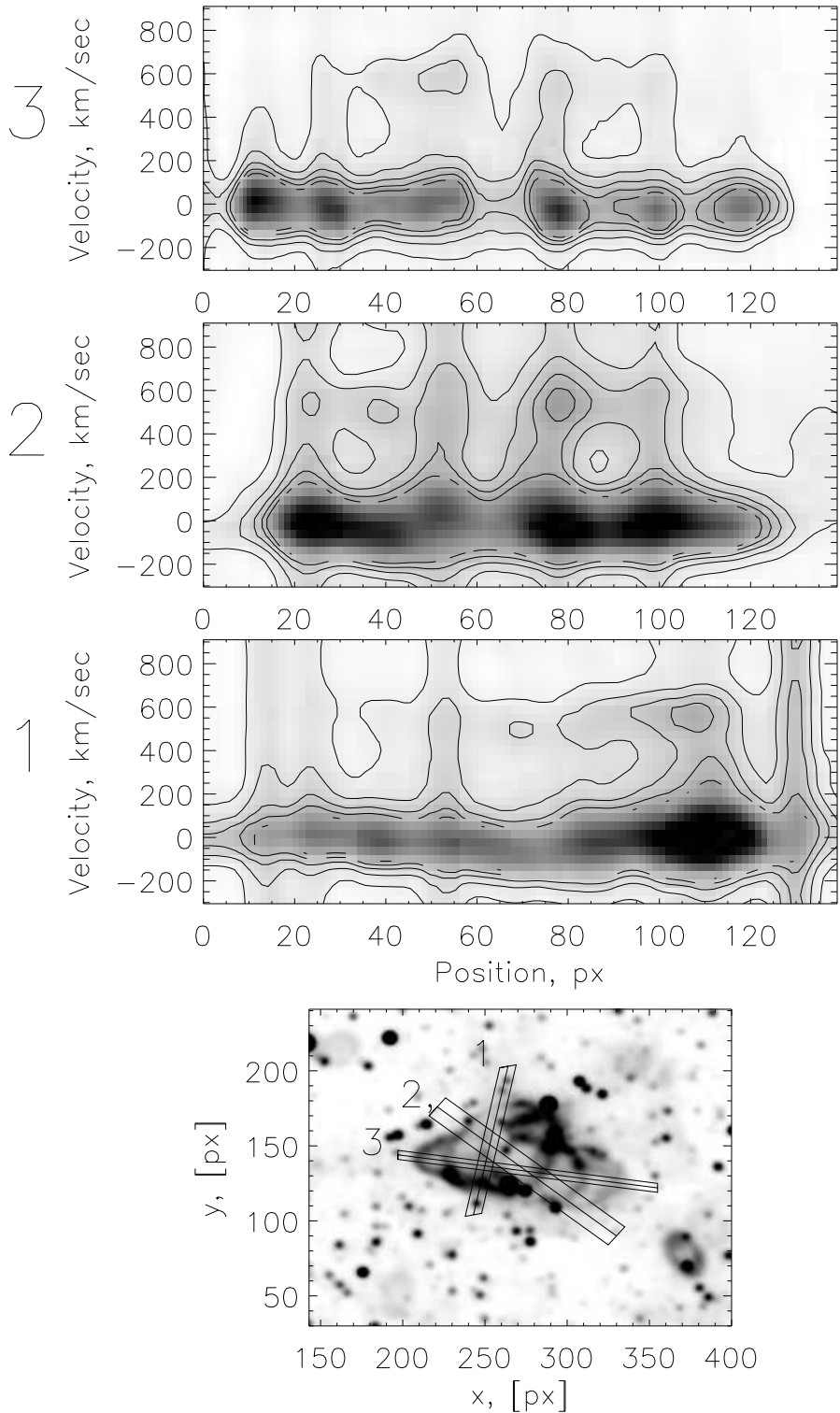}
\caption{Position--velocity diagrams constructed from our FPI observations
and the localization of the corresponding scans in the PWN image.
The emission features at a velocity of about 600~km~s$^{-1}$ are attributable
to the [NII] 6548\,\AA\ line. \hfill}
\end{figure*}

The filamentary morphology and the intense [SII] line emission provide strong
evidence for the gas emission behind the shock front. It is interesting to
measure the velocities of these filaments in the extended remnant outside the
core, since, in general, the thin filaments can be localized on the front or
rear side of the old shell; i.e., these may be projected rather than be
physically associated with the PWN. Besides, the pulsar's velocity with
respect to the ambient gas is important in estimating the expected shape
of the bow shock produced by the wind from a fast-moving pulsar. Figure~4
(bottom panel) shows only the brightest outer filaments in a $4'\times4'$
field for which the background subtraction effect is insignificant (see
above). The H$\alpha$ brightness of these filaments exceeds
$8.6\times 10^{-17}$~erg~s$^{-1}/\sq^{''}$. As follows from Fig.~4 (bottom),
the velocities of the bright filaments in the outer shell of CTB\,80 do not
differ significantly from the velocities of the peripheral filaments of
the PWN undistorted by the expansion of the latter.

\begin{figure}[t!]
\vspace*{0.1cm}
\includegraphics[width=8.6cm,bb=175 120 435 675,clip]{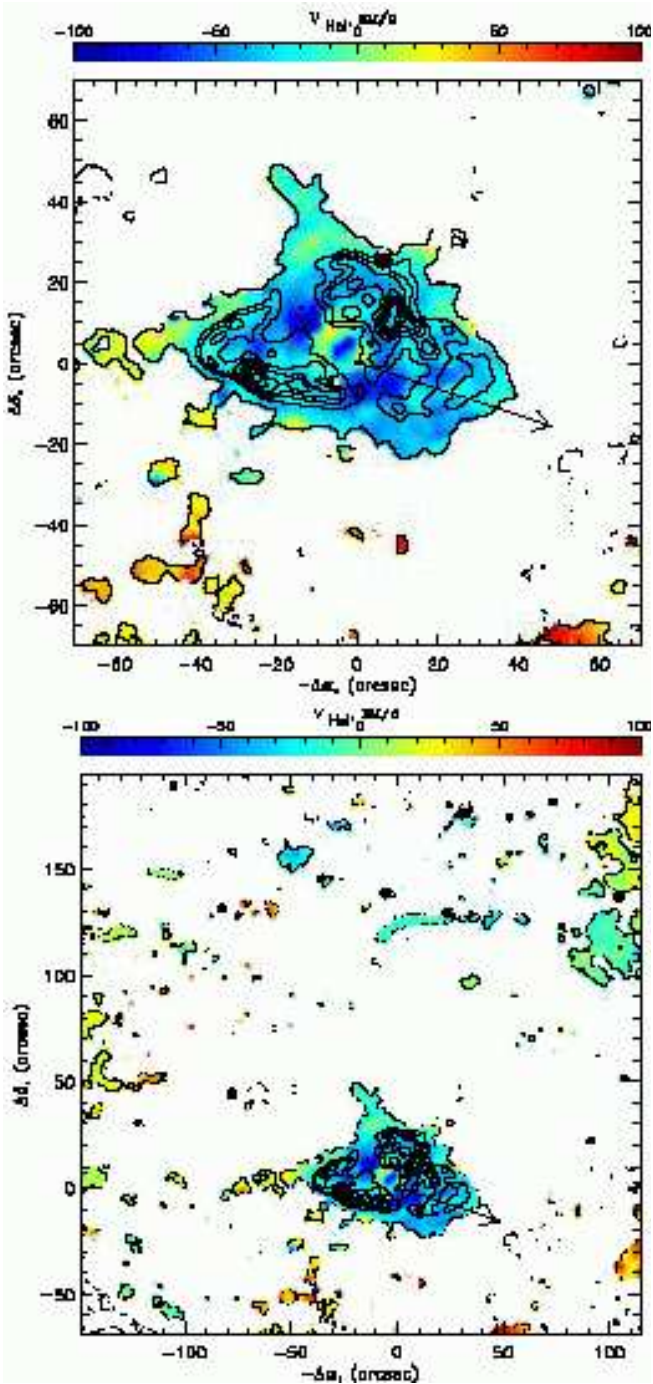}
\caption{CTB80 core. {\it Top panel}: the velocity field of the H$\alpha$
peak superimposed on the PWN image. The arrow indicates the direction of
the pulsar's motion, as measured by Migliazzo \etal\,(2002); {\it bottom
panel}: the same for the entire FOV of about~$5' \times 5'$ in size in
the image obtained with the 6\,m telescope (only the brightest outer
filaments are shown). \hfill}
\end{figure}

Note, in particular, the filaments that are immediately adjacent to the PWN
in the north and the east and that form a kind of a wake at the boundary of
the shell produced by the wind from a moving pulsar. (These filaments are
most distinct in the [SII] lines in  Fig.~2b from Hester and Kulkarni~1989).
As our measurements indicate, the velocities of these filaments do not differ
significantly from those of the bright PWN filaments either.

\section*{DISCUSSION}

The difficulty of explaining the nature of the PWN in CTB\,80 lies in the
fact the bow-shock structure (in the [OIII] line, the radio continuum, and
the soft X-ray band) and the elongated filamentary shell-like structure in
the H$\alpha$ line, which is definitely also of an shock origin that, in
the opinion of Hester~(2003), is associated with the action of the pulsar's
jets, are simultaneously clearly seen around the pulsar. We do not see any
manifestations of the interaction between two processes: the ejections
of relativistic pulsar plasma in two jets and the motion of the pulsar.
The elongated PWN structure is not distorted by the pulsar's motion in any
way; the bow shock, in turn, is not distorted by the action of the two plasma
jets in any way either. The rough estimates given below indicate that
individually both structures can be explained adequately.

\subsection*{Estimates for the Shock Waves Produced by the Pulsar's Jets}

If the pulsar's power~$L$ is constant over time~$t$ and is released in
a solid angle~$\Omega$, blowing a bubble of radius~$R=R(t)$ in a homogeneous
interstellar medium with a density~$\rho_0$, then the mass $ M \sim \Omega
\rho_0 R^3 /3$ is swept up from the bubble. Assuming that all of this mass is
gathered in a thin layer at radius~$R$ and remains in the same solid
angle~$\Omega$ (in fact, it will partially flow around the jet) and equating
the pulsar-wind momentum flux~$L/c$ to the rate of increase in the
momentum of the wind-driven shell,
$$
\frac{L}{c}=\frac{d}{dt}\left(Mv\right),
$$
where $v=dR/dt$, we obtain an acceptable size of the bubble swept up by
the wind for a pulsar at rest under reasonable assumptions about the ambient
density and the wind asphericity. Assuming that $R \propto t^\alpha$, we have
$$
\frac{L}{c}=\frac{\Omega}{3} \rho_o \alpha
                \frac{d}{dt}\left(\frac{R^4}{t}\right);
$$
hence, $\alpha=1/2$ at $L={\textrm{const}}$, i.e.,
\begin{equation}
\label{rwind:Lozinskaya_n}
\hspace*{0.25cm} R = \left(\frac{6L}{\Omega c \rho_o}\right)^{1/4} t^{1/2}
\sim 3 \mbox{~pc}\times\left(\ell L_{36} \frac{4\pi}{\Omega n_0}\right)^{1/4} t_6^{1/2},
\end{equation}
where $ \ell=L/L_0$ is the fraction of the power~$L_0$ went into the solid
angle~$\Omega$ and spent on blowing the bubble, $n_0$ is the particle number
density (for hydrogen composition), and~$t_6$ is the time in Myr.

The power of the flux in relativistic particles (including all types of
photons) from the pulsar PSR~B1951$+$32 is $ L_0=\dot E_{\textrm{rot}} = 3.7
\times 10^{36}$~erg~s$^{-1}$ (Kulkarni \etal\,1988). At $n_0=1$~cm$^{-3}$ ,
$L_{36}=3.7$, and the pulsar's total age $t_6=0.1$, we obtain a bubble size
of about 1~pc at $\ell =1 $ even for isotropic radiation ($\Omega =4\pi$), in
agreement with the observed PWN size in the H$\alpha$ line. In fact, only
a small fraction~$\ell$ of the pulsar's power goes into blowing the bubble,
but~$\Omega$ is also much less than~$4\pi$ (by several hundred times for
a linear jet opening angle of several degrees).

We emphasize that this is the minimum upper limit: here, the shock waves are
assumed to be radiative. The bubble size will increase if the hot-gas pressure
inside the bubble is taken into account (Castelleti \etal\,2003). Two cases
are possible: the radius for radiative shock waves is larger than our
estimate, but smaller than that for adiabatic shock waves (McKee and
Ostriker~1977; Blinnikov \etal\,1982).

If we took into account the hot-gas pressure inside the bubble and for
adiabatic shock waves, we would obtain the maximum estimate at $\Omega =4\pi$
(Avedisova~1971; Weaver \etal\,1977; Ostriker and McKee~1988):
\begin{equation}
\label{rweav:Lozinskaya_n}
\hspace*{2.5cm} R=28 $~pc$\times n_0^{-1/5} L_{36}^{1/5} t_6^{3/5} .
\end{equation}
Hence, $R$ could be even an order of magnitude larger than that from~(1).

Let us make several remarks suggesting that these rough estimates are
uncertain. On the one hand, the Balmer line emission in PWNs is evidence of
nonradiative shock waves in a partially neutral medium (Chevalier and
Raymond~1980); it is produced by electron impact excitation or ion charge
exchange. In this case, neutral hydrogen can penetrate into the hot gas and
impart odd shapes to the shock waves (Bucciantini and Bandiera~2001;
Bucciantini~2002; D'Amico \etal\,2003)

On the other hand, formula~(2) was derived for a hot wind with a normal
(Pascal) pressure. Since the flow of
magnetized particles or photons in a pulsar wind is directed
only along the radius, a simple formula of form~(1) the derivation
of which assumes no pressure isotropy holds good, while formula~(2)
overestimates the result. Many problems associated with the composition of
this wind have not yet been solved (the so-called
sigma paradox --- the transition from the fraction of the electromagnetic
pressure expressed in terms of the Poynting vector to the kinetic pressure
of the particle flux cannot be reliably calculated (see the review by
D'Amico \etal\,2003)).

A time shorter than the pulsar's age should be taken as~$t_6$, since it has
flown into the dense layers of the shell of the old SNR relatively recently.
Castelleti \etal\,(2003) took $t=18\,200$\,yr as the PWN age at an ambient
density of $n_0 = 0.5$~cm$^{-3}$. According to Mavromatakis \etal\,(2001),
the relative line intensities in the spectrum of the filaments of the
extended CTB\,80 shell are typical of the gas radiation behind the front of
a shock propagating at a velocity of 85--120~km~s$^{-1}$ in a medium with
an initial density of about 2--5~cm$^{-3}$. This density is equal to the mean
density in the HI shell estimated by Koo \etal\,(1990).

However, the dependence of~$R$ on all parameters, including the density and
the age, is weak; i.e., clearly, a small fraction~$\ell$ of the total power
that goes into blowing the bubble will suffice. (We also see the pulsar's
emission; i.e., clearly, the fraction of the captured power, $\ell$, is
appreciably smaller than unity.)

Thus, the sizes of the bubbles swept up by the pulsar-wind jets can easily
be made close to the observed values even if the density is much higher and
the fraction~$\ell$ of the pulsar's power that goes into the jet is small.

\subsection*{An Estimate for the Location of the Bow Shock}

The pulsar's motion in the plane of the sky with a velocity of
$v_0=240$~km~s$^{-1}$ (Migliazzo \etal\,2002) for an \emph{isotropic} wind
must give rise to a bow shock, which is observed in several pulsars (see
the review by D'Amico \etal\,2003). To estimate the shock shape, we can use
the solutions by Lipunov and Prokhorov~(1984) and Wilkin~(1996). The latter
author obtained an analytical solution for a normal (nonrelativistic) stellar
wind with radiative shock waves:
\begin{equation}
\hspace*{2.0cm} R(\theta)=d_0\csc\left[3(1-\theta \cot \theta)
\right]^{1/2},
\label{wilkin:Lozinskaya_n}
\end{equation}
where $d_0$ is the distance from the pulsar to the head point at contact
discontinuity. This solution describes well the classical numerical results
by Baranov \etal\,(1971) and van der Swaluw \etal\,(2003), see Fig.8 in the
latter. An approximate solution for adiabatic shock waves was given by Chen
\etal\,(1996). The criticism of the applicability of such solutions to known
PWNs (see Bucciantini and Bandiera~2001) is based on the fact that these do
not include the effects of charge exchange and neutral hydrogen penetration
behind the shock mentioned above. However, the PWN in CTB\,80 is unique in
that the bow shock is observed not only in Balmer lines, but also, most
clearly, in [OIII] lines and in the X-ray and radio bands, i.e., in a hot,
ionized plasma. Therefore, simple analytical solutions may well be valid
in this case.

\begin{figure}[t!]
\vspace*{0.05cm}
\includegraphics[width=8.6cm,bb=172 262 420 675,clip]{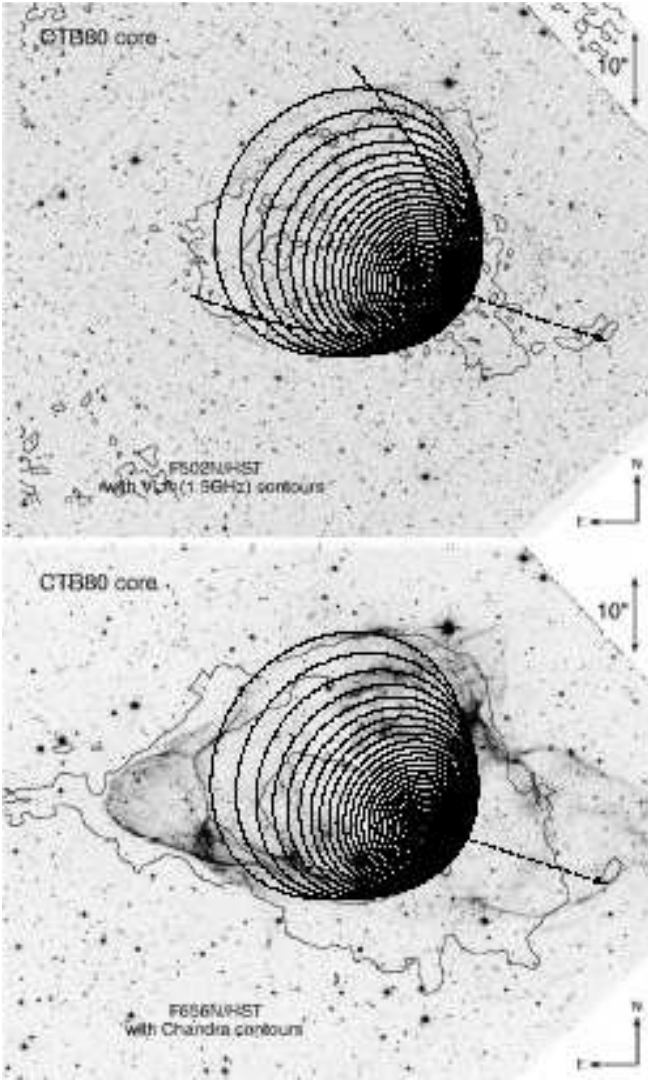}
\caption{{\it Top panel}: PWN in the [OIII] line and in the radio continuum
(isophotes): the bow shock corresponding to solution~(3) for the pulsar's
motion in the plane of the sky with a position angle of $235^\circ$ (dashed
line) and the bow shock inclined at an angle of $60^{\circ}$ to the plane of
the sky ($\theta$ isolines). {\it Bottom panel}: The same bow shock inclined
at an angle of $60^\circ$ chosen by the best agreement with the [OIII] line
image superimposed onto the H$\alpha$ and X-ray image (isophotes) of the PWN.
The arrow indicates the pulsar's motion over a period of 1000\,yr, as measured
by Migliazzo et al. (2002). \hfill}
\end{figure}

We obtain the following estimate from the balance between the momentum fluxes
$\ell_1 L/c$ and $4\pi d_0^2 \rho_o v_0^2$:
\begin{equation}
\label{d0:Lozinskaya_n}
\hspace*{3.0cm} d_0=\left(\frac{\ell_1 L}{4\pi c \rho_o  v_0^2}
\right)^{1/2}.
\end{equation}
Hence, we find for $v_0=240$~km~s$^{-1}$, $\rho_o=2\times 10^{-24}$~g~cm$^{-3}$
($n_0=1$~cm$^{-3}$), and $\ell_1=1$ that $d_0 \approx 0.044$~pc. This value is
comparable to the observed distance from the pulsar to the [OIII] filaments
that determine the position of the head point at the shock front,
$d_{\textrm{obs}}\sim 6'' \approx 0.057$~pc.

The theoretical value of~$d_0$ will be close to~$d_{\textrm{obs}}$ if we take
into account the fact that $v_0=170$~km~s$^{-1}$ should be substituted for
$v_0=240$~km~s$^{-1}$, since the pulsar moves in the matter of a remnant
expanding with a velocity of 72~km~s$^{-1}$. In this case, $d_0 \approx
0.043$~pc.

In Fig.\,5 (top panel), the solution by Wilkin~(1996) is superimposed on
the PWN image in the [OIII] line and in the radio continuum. The distance
from the head point at the front to the pulsar is assumed to be $d_0 \approx
0.043$~pc, and the axis of the surface with a position angle of $235^\circ$
does not coincide with the pulsar's velocity direction in the plane of
the sky. An azimuthal asymmetry in the projection of the theoretical surface
relative to the observed shape of the central shell remains noticeable at
a position angle of $P=252^\circ$ according to the measurements by Migliazzo
\etal\,(2002).

The fact that the shape of the bow shock observed in the [OIII] line is
appreciably ``broader'' than the gas-dynamical solutions by Wilkin~(1996) may
suggest that the pulsar moves at a significant angle to the plane of the sky.
Allowing for the possible inclination of the pulsar's velocity vector to
the plane of the sky yields better agreement of the theory with the observed
morphology of the central shell. By varying the pulsar's velocity vector, we
found the best agreement with the observations for a bow shock whose axis is
inclined at an angle of $60^\circ$ to the plane of the sky.
Figure~5 (top panel) shows this solution for the bow shock chosen by
the best agreement with the shape of the central shell observed in the [OIII]
line and in the radio continuum. As we see, the shape of the projection of
the theoretical surface defined by relation~(3) can be completely reconciled
with the observed PWN morphology. In this case, the position angle of the
pulsar's velocity vector in the plane of the sky is $P=235^\circ$; i.e., it
is close to the direction of the symmetry axis in the velocity field of the
main H$\alpha$ line component found above. $P=235^\circ$ matches the value
obtained by Strom~(1987).

We emphasize that, if the pulsar's velocity vector is inclined at an angle of
$60^\circ$, its space velocity is twice as high as its velocity measured in
the plane of the sky; i.e., it reaches 500~km~s$^{-1}$ (this value agrees
with the mean value in the velocity distribution of pulsars; see Arzoumanian
\etal\,2002.) Since the distance~$d_0$ at a higher velocity is a factor
of~$\sqrt{2}$ smaller, for close agreement with the observations, we must
take a slightly lower gas density in front of the bow shock.

Our kinematic studies are also consistent with the suggested model. Since
the thickness of the postshock emitting gas, which is determined by
the apparent thickness of the bright peripheral PWN filaments, is an order
of magnitude smaller than the distance the pulsar traverses in 1000\,yr, (see
Fig.~1), we conclude that the radiative cooling time of the postshock gas is
short, about 100\,yr. In this time, an element of gas is not only
compressed, but also acquires a velocity that corresponds to the pulsar's
motion. Without detailed calculations, it is hard to tell at which velocity
the gas emission is at a maximum; one may only expect the observed velocity
to be a significant fraction of the pulsar's velocity.

Gas motions with such velocities in the PWN have been detected in our work
for the first time (see above). These velocities refer to the weak emission
features in the H$\alpha$ line; they are observed both in the central region
around the pulsar, and near the bright filaments at the PWN boundary (see
Fig.~2). The line-of-sight velocities of the bright filaments are within
the $\pm 200$~km~s$^{-1}$ range. This is an ordinary (for SNRs) situation
related to the fact that the bright filaments usually represent the front
surfaces seen edge-on.

Unfortunately, since the velocity measurement range in our FPI observations
is limited, at present, we cannot unambiguously determine whether the pulsar
is moving toward or away from us. The existence of high positive velocities
argues for the motion away from us, but as yet we have no information about
the possible negative velocities without further FPI observations. The derived
velocity distribution of the line peak could clarify the picture, but here we
are restricted by the fact that the density distribution in the closest
neighborhood of the shell corresponding to the bow shock is unknown. In
particular, the asymmetric expansion of the central shell (according to
Whitehead \etal\,(1989), the motion of the approaching side at a velocity of
about $-200$~km~s$^{-1}$ is most clearly observed in it) could be attributable
to the higher brightness of this side due to a nonuniform ambient gas density.

In Fig.~5 (bottom panel), the same theoretical surface of the bow shock that
is inclined at an angle of $60^\circ$ to the plane of the sky and that agrees
best with the PWN emission in the [OIII] line is superimposed on the H$\alpha$
image with X-ray isophotes. The two shell-like H$\alpha$~structures in the
west and the east are far outside the bow shock. These structures, which form
the elongated PWN shape, were explained by the action of the pulsar's jets
(Hester~2000).

It should be noted that, if the pulsar's jets are directed at an angle to its
space velocity (e.g., lie in the plane of the sky), then the contradiction
with the absence of an apparent influence of the jets on the shock front in
its brightest parts mentioned at the beginning of the section is removed.

Of course, the representation of the shock as an infinitely thin layer with
shape~(3) is oversimplified. More detailed solutions should be used, and
the structure of two shocks with a contact discontinuity between them should
be taken into account (Baranov \etal\,1976), since there is a hint at such
a structure in the observations. In particular, Moon \etal\,(2004) associate
the structure observed in X~rays with a backward shock wave in the pulsar
wind and in the H$\alpha$ line with a bow shock in the ambient medium.

Note also that, in fact, $\ell_1$ in formula~(4) may be the fraction of
the pulsar's emission absorbed by an ionized plasma, while~$\ell$ in
formula~(1) for the jet may be the fraction of the emission absorbed by
neutral hydrogen. For example, hard ultraviolet emission must be absorbed
completely by neutral hydrogen and only partially by ionized plasma. Another
possible process, the proton-beam charge exchange in a neutral gas, cannot
take place in an ionized medium. Thus, different components of the pulsar's
flux can manifest themselves differently in different components of the
ambient medium.

\begin{figure*}[t!]
\hspace*{0.7cm}
\includegraphics[width=17.0cm,bb=55 190 555 600,clip]{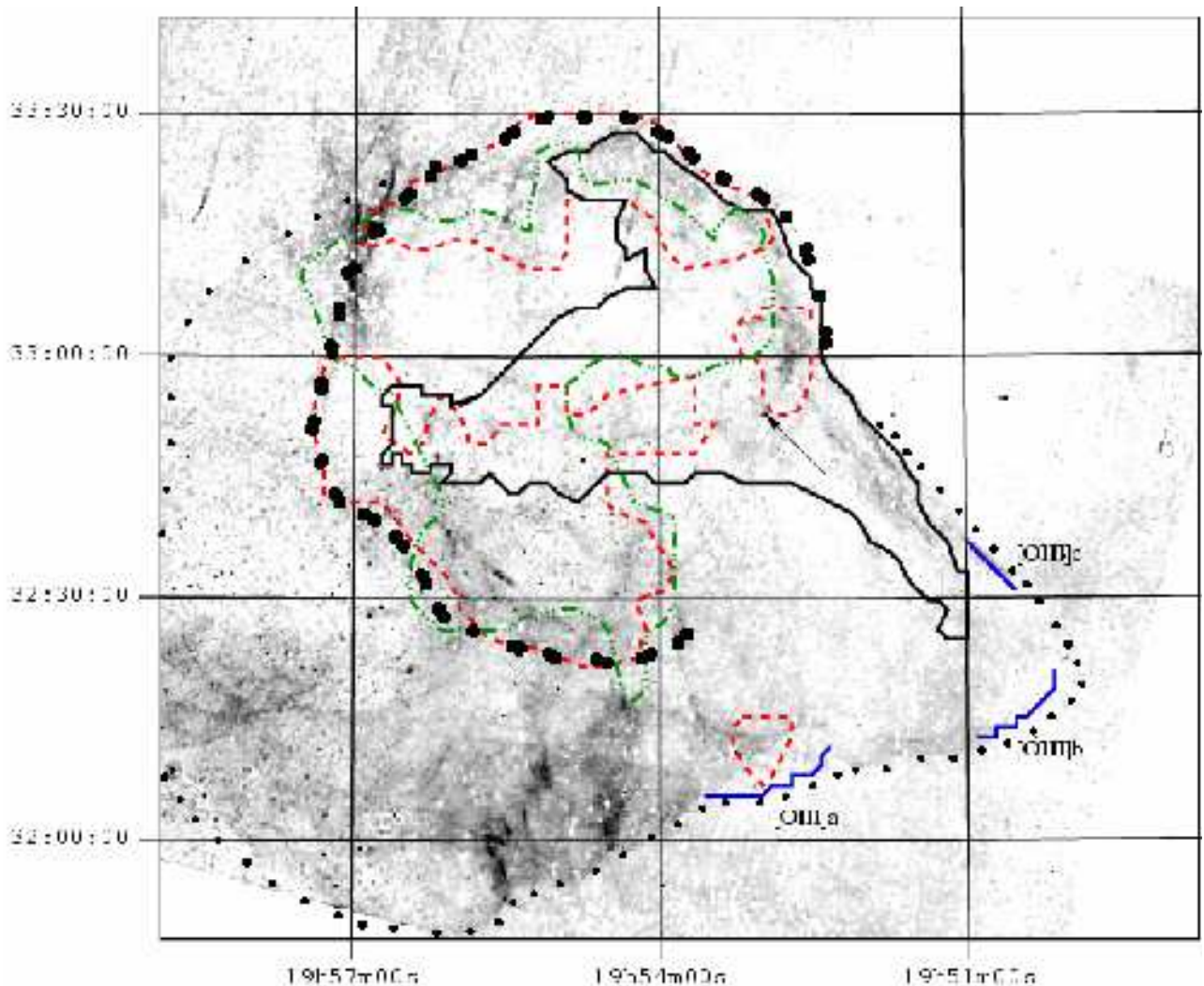}
\caption{General scheme of CTB\,80 from Mavromatakis \etal\,(2001): the [SII]
image of the region, radio ridges (solid line), and the infrared and HI~shells
(the dashed and dash--dotted lines, respectively). The large and small circles
indicate the suggested model for the shell of the old SNR composed of two
hemispheres in projection onto the plane of the sky. \hfill}
\end{figure*}

A different interpretation of the elongated PWN shape in the H$\alpha$ line
is also possible. The system of thin filaments mentioned in the Section
``Results of Observations'', which characterizes the optical emission from
the extended remnant outside the PWN, is most likely the layered structure
of the old CTB\,80 shell (Hester and Kulkarni~1989). Our velocity measurements
for the brightest filaments suggest that these are not projected, but are
physically associated with the PWN.

The thickness of the layers (typically, $\sim 10\text{--}20''$ or 0.1--0.2~pc)
and the separation between them (about $2'\text{--}2.5'$ or 1--1.5~pc) at
the pulsar's velocity of 240~km~s$^{-1}$ in the plane of the sky yield a time
of about 5000\,yr between the passages of neighboring layers and a passage
time of a dense layer of 400--800\,yr.

The pulsar's motion through such a layered medium can produce the observed
elongated multishell structure of the core as a result of the pulsar-wind
breakthrough from a dense layer into a tenuous medium between the layers.
The breakthrough of the nonrelativistic wind of Wolf--Rayet stars from a
dense cloud into a tenuous intercloud gas leads precisely to this effect
(see, e.g., Dopita and Lozinskaya~1990).

\subsection*{Localization of the Pulsar in the Extended Remnant CTB\,80}

If the space velocity of the pulsar actually reaches 500 km~s$^{-1}$, then,
in general, it traverses a distance of 50~pc in a time of $10^5$\,yr and
could go beyond the symmetric infrared and H\,I shells about 40~pc in size,
which, according to the previous interpretation, determined the total volume
of the old remnant CTB\,80. However, the deep images taken in
the H$\alpha+$[NII], [SII], [OII], [OIII] lines by Mavromatakis \etal\,(2001)
revealed large-scale filamentary and diffuse structures in the nearby
$2^{\circ}\times2^\circ$ region. These optical filaments in the north and
the northwest closely correlate with the radio images of the remnant
(Castelleti \etal\,2003) and with the boundary of the infrared shell, but
go far beyond the infrared and H\,I shells in the south and the east. (Here,
we disregard the extended filaments in the east associated not with CTB\,80,
but with the remnant of a different supernova (Mavromatakis and Strom~2002).)
The relative line intensities in the spectrum of these large-scale filaments
are typical of the radiative cooling of the gas behind the front of a shock
propagating at a velocity of 85\mbox{--}120~km~s$^{-1}$ in a medium with an
initial density of about 2--5~cm$^{-3}$ (Mavromatakis \etal\,2001). Such a
density agrees with the mean density in the H\,I shell estimated by
Koo \etal\,(1990) from 21-cm line observations. This argues for the
association of the large-scale filaments with the SNR CTB\,80.

Therefore, a more complex spatial structure composed of two hemispheres
with different sizes should probably be considered as the SNR CTB\,80.
One hemisphere, which is determined by the northeastern radio ridge and
the infrared and H\,I shells, is the result of the interaction of the shock
triggered by a supernova explosion with a dense medium. The second part of
the shell has an aspherical shape and is determined by the southwestern ridge
and the [OIII] filaments denoted by~III and~IV as well as by the system of
bright filaments to the east of~IV (see Fig.~2 from Mavromatakis \etal\,2001).
This part of the shell was most likely formed by the shock in a medium with
a much lower density.

Figure~6 shows approximate boundaries of the two parts of the shell in
projection onto the plane of the sky in this model.

Note that ROSAT observations revealed a conical $\sim 1^\circ$ region of
thermal X-ray emission southeast of the pulsar far beyond the infrared shell
(Safi-Harb \etal\,1995). The central region of the second part of the shell
mentioned above could, in principle, be responsible for this emission.
However, the SNR CTB\,80 is observed along the Cygnus spiral arm and is
immediately adjacent to the giant Superbubble produced by intense stellar
wind from the Cyg~OB2 cluster (Lozinskaya \etal\,2002, and references
therein). Therefore, in general, the thermal X-ray emission in the extended
conical region could be the background emission, i.e., it could belong not to
CTB\,80, but to the Superbubble.

In the proposed scheme, the morphology of the SNR in the plane of the sky
suggests that the major axis of this structure composed of two hemispheres
with different sizes is oriented at a large angle to the plane of the sky.
Therefore, at a possible space velocity of 500~km~s$^{-1}$, the pulsar has
not yet gone outside the shell of the extended remnant CTB\,80.

\section*{CONCLUSIONS}

Our kinematic study of the pulsar wind nebula in the old supernova remnant
CTB\,80 using the FPI of the 6\,m SAO telescope has revealed weak
high-velocity H$\alpha$ features in the PWN at least up to a velocity of
400--450~km~s$^{-1}$. We confirmed the previously measured expansion of
the system of bright filaments with a velocity of 100--200~km~s$^{-1}$.
We analyzed the PWN morphology in the H$\alpha$, [SII], and [OIII] lines
using the HST archival data. The shape of the bow shock, which is
determined by the central horseshoe-shaped shell bright in the [OIII] line
and in the radio continuum, was shown to be in best agreement with the theory
for a significant (about~60$^\circ$) inclination of the pulsar's velocity
vector to the plane of the sky. The space velocity of the pulsar is twice as
high as its tangential velocity measured by Migliazzo \etal\,(2002); i.e.,
it reaches 500~km~s$^{-1}$. This pattern of motion is also confirmed
by the high radial velocities of the gas in the PWN that we found here. Thus,
PSR~B1951$+$32 is the first pulsar whose possible light-of-sight velocity
(of about 400 km s$^-1$) has been estimated from the PWN morphology and
kinematics.

The filamentary shell-like structures observed in the H$\alpha$ line in
the east and the west outside the bow shock can be explained not only by
the action of the pulsar's jets, but also by the pulsar-wind breakthrough
into an inhomogeneous ambient medium.

We considered the general scheme of~CTB\,80 that includes the most recent
optical and radio observational data in which the pulsar's high space
velocity is consistent with its location in the dense shell of the old SNR.

Of course, the proposed scheme of the PWN in CTB\,80 must be confirmed
additionally. The existence of high velocities in the PWN, the pattern of
elongated H$\alpha$~structures (possibly under the influence of the pulsar's
jets or when the wind breaks through from a thin dense layer), and other
questions require further observations and detailed hydrodynamic simulations
of this interesting object. An analysis of the velocity field for the extended
filaments in the entire region can give a kinematic confirmation for the
proposed scheme of the old SNR CTB\,80.

\section{ACKNOWLEDGMENTS}

This work was supported by the Russian Foundation for Basic Research (project
nos.~02-02-16500, 03-02-17423, \hspace{-2pt}04\mbox{-}02\mbox{-}16042)
\hspace{-1pt} and the Federal Science and Technology Program\hspace{-1pt}
(contract no.~40.022.1.1.1103). A.V.~Moiseev thanks the Russian Sciences
Support Foundation for partial support of this work. We are grateful to
Yu.A.~Shibanov for a kind permission to use the data
obtained as part of his observational program on the 6\,m BTA telescope,
B.M.~Gaensler for providing the VLA radio data and I.V.~Karamyan,
V.Yu.~Avdeev, and O.V.~Egorov for help. This work is based on the
observational data obtained with the 6\,m SAO telescope financed by
the Ministry of Science of Russia (registration no.~01-43) and on
the NASA/ESA Hubble Space Telescope data taken from the archive of
the Space Telescope Science Institute operated by the Association of
Universities for research in astronomy under a NASA contract (NAS~5-26555)
and data of the Chandra X-ray Observatory Center, which is operated by
the Smithsonian Astrophysical Observatory for and on behalf of the National
Aeronautics Space Administration under contract NAS8-03060.

Translated by V.~Astakhov


\begin{thebibliography}{99}

\bibitem{5:Lozinskaya_n}
V.~L.~Afanasiev and A.~V.~Moiseev, Pis'ma Astron.Zh. \textbf{31}, 214 (2005),
[Astron. Letters, 31, 193, (2005)]

\bibitem{3:Lozinskaya_n}
P.~E.~Angerhofer, R.~G.~Strom, T.~Velusami, \emph{et al.} Astron. Astrophys.
\textbf{94}, 313 (1981).

\bibitem{2:Lozinskaya_n}
P.~E.~Angerhofer, A.~S.~Wilson, and J.~R.~Mould, Astrophys. J. \textbf{236}, 143 (1980).


\bibitem{4:Lozinskaya_n}
Z.~Arzoumanian, D.F.~Chernoff, and J.M.~Cordes, Astrophys. J. \textbf{568}, 289 (2002).

\bibitem{1:Lozinskaya_n}
V.~S.~Avedisova, Astron. Zh. \textbf{48}, 894 (1971) [Sov. Astron. \textbf{15}, 708
(1972)].

\bibitem{6:Lozinskaya_n}
V.~B.~Baranov, K.~V.~Krasnobaev, and A.~G.~Kulikovski@i, Dokl. Akad. Nauk SSSR
\textbf{194}, 41 (1971) [Sov. Phys.--Dokl. \textbf{15}, 791 (1971)].

\bibitem{7:Lozinskaya_n}
V.~B.~Baranov, K.~V.~Krasnobaev, and M.~S.~Ruderman, Astrophys. Space Sci. \textbf{41},
481 (1976).

\bibitem{10:Lozinskaya_n}
S.~I.~Blinnikov, V.~S.~Imshennik, and V.~P.~Utrobin, Pis'ma Astron. Zh. \textbf{8}, 671
(1982) [Sov. Astron. Lett. \textbf{8}, 361 (1982)].

\bibitem{9:Lozinskaya_n}
W.~P.~Blair, R.~A.~Fesen, and R.~H.~Becker, Astron. J. \textbf{96}, 1011 (1988).

\bibitem{8:Lozinskaya_n}
W.~P.~Blair, R.~P.~Kirshner, R.~A.~Fesen, \emph{et al.}, Astrophys. J. \textbf{282}, 161
(1984).

\bibitem{11:Lozinskaya_n}
N.~Bucciantini, Astron. Astrophys. \textbf{387}, 1066 (2002).

\bibitem{12:Lozinskaya_n}
N.~Bucciantini, Astron. Astrophys. \textbf{393}, 629 (2002).

\bibitem{13:Lozinskaya_n}
N.~Bucciantini and R.~Bandiera, Astron. Astrophys. \textbf{375}, 1032 (2001).

\bibitem{18:Lozinskaya_n}
G.~Castelleti, G.~Dubner, K.~Golap, \emph{et al.}, Astron. J. \textbf{126}, 2114 (2003).

\bibitem{46:Lozinskaya_n}
Y.~Chen, R.~Bandiera, and Z.~Wang, Astrophys. J. \textbf{469}, 715 (1996).

\bibitem{47:Lozinskaya_n}
R.~A.~Chevalier and J.~C.~Raymond), Astrophys. J. \textbf{225}, L27 (1980).

\bibitem{16:Lozinskaya_n}
N.~D'Amico, R.~Bandiera, N.~Bucciantini, \emph{et al.}, Mem. Soc. Astron. Italiana
\textbf{74}, 345 (2003).

\bibitem{17:Lozinskaya_n}
M.A.~Dopita and T.A.~Lozinskaya, Astrophys. J. \textbf{359}, 419 (1990).

\bibitem{40:Lozinskaya_n}
R.~A.~Fesen, J.~M.~Shull, and J.~M.~Saken, Nature \textbf{334}, 229 (1988).

\bibitem{41:Lozinskaya_n}
A.S.~Fruchter, J.H.~Taylor, D.C.~Backer, \emph{et al.}, Nature \textbf{331}, 53 (1988).

\bibitem{42:Lozinskaya_n}
J.~J.~Hester, \emph{Proc. Conf.``Spin, Magnetism and Cooling of Young Neutron Stars'',
Kavli Institute of Theoretical Physics, 2000},
http://online.itp.ucsb.edu/online/neustars\_c00/hes\-ter/.

\bibitem{43:Lozinskaya_n}
J.~J.~Hester, Bull. Am. Astron. Soc. \textbf{32}, 1542 (2000).

\bibitem{44:Lozinskaya_n}
J.~J.~Hester and S.~R.~Kulkarni, Astrophys. J. \textbf{331}, L121 (1988).

\bibitem{45:Lozinskaya_n}
J.~J.~Hester and S.~R.~Kulkarni, Astrophys. J. \textbf{340}, 362 (1989).

\bibitem{19:Lozinskaya_n}
B.-C.~Koo, W.T.~Reich, C.~Heiles, \emph{et al.}, Astrophys. J. \textbf{364}, 178 (1990).

\bibitem{20:Lozinskaya_n}
B.~C.~Koo, M.~S.~Yun, P.~T.~P.~Ho, \emph{et al.}, Astrophys. J. \textbf{417}, 196 (1993).

\bibitem{21:Lozinskaya_n}
S.~R.~Kulkarni, T.~C.~Clifton, D.~C.~Backer, \emph{et al.}, Nature \textbf{331}, 50
(1988).

\bibitem{22:Lozinskaya_n}
V.M.~Lipunov and M.E.~Prokhorov, Astrophys. Space Sci. \textbf{98}, 221 (1984).

\bibitem{23:Lozinskaya_n}
T.~A.~Lozinskaya, V.~V.~Pravdikova, and A.~V.~Finogenov, Pis'ma Astron. Zh. \textbf{28},
260 (2002) [Astron. Lett. \textbf{28}, 223 (2002)].

\bibitem{27:Lozinskaya_n}
F.~Mantovani, M.~Reich, C.~J.~Salter, \emph{et al.}, Astron. Astrophys. \textbf{145}, 50
(1985).

\bibitem{24:Lozinskaya_n}
F.~Mavromatakis and R.~G.~Strom, Astron. Astrophys. \textbf{382} 291, (2002).

\bibitem{25:Lozinskaya_n}
F.~Mavromatakis, J.~Ventura, E.~V.~Paleologou, \emph{et al.}, Astron. Astrophys.
\textbf{371}, 300 (2001).

\bibitem{26:Lozinskaya_n}
C.~F.~McKee and J.~P.~Ostriker, Astrophys. J. \textbf{218}, 148 (1977).

\bibitem{28:Lozinskaya_n}
J.~M.~Migliazzo, B.~M.~Gaensler, D.~C.~Backer, \emph{et al.}, Astrophys. J.
\textbf{567}, L141 (2002).

\bibitem{29:Lozinskaya_n}
A.~V.~Moiseev, Bull. Spec. Astrophys. Obs. \textbf{54}, 74 (2002); astro-ph/0211104.

\bibitem{30:Lozinskaya_n}
D.-S.~Moon, J.-J.~Lee, S.~S.~Eikenberry, \emph{et al.}, Astrophys. J. \textbf{610}, L33
(2004).

\bibitem{31:Lozinskaya_n}
J.~P.~Ostriker and C.~F.~McKee, Rev. Mod. Phys. \textbf{60}, 1 (1988).

\bibitem{32:Lozinskaya_n}
S.~Safi-Harb, H.~Ogelman, and J.~P.~Finley, Astrophys. J. \textbf{439}, 722 (1995).

\bibitem{33:Lozinskaya_n}
R.~G.~Strom, Astrophys. J. \textbf{319}, L103 (1987).

\bibitem{36:Lozinskaya_n}
R.~G.~Strom, P.~E.~Angerhofer, and J.~R.~Dickel, Astron. Astrophys. \textbf{139}, 43
(1984).

\bibitem{34:Lozinskaya_n}
R.~G.~Strom and W.~P.~Blair, Astron. Astrophys. \textbf{149}, 259 (1985).

\bibitem{35:Lozinskaya_n}
R.~G.~Strom and B.~W.~Stappers, \emph{Proc. Conf. ''Pulsar Astronomy''} (2000).

\bibitem{}
E. van der Swaluw, A. Achterberg, Y. A. Gallant,
T. P. Downes, R. Keppens, Astron.Astrophys. 397, 913 (2003).

\bibitem{14:Lozinskaya_n}
T.~Velusami and M.~R.~Kundu, Astron. Astrophys. \textbf{32}, 375 (1974).

\bibitem{15:Lozinskaya_n}
T.~Velusami, M.~R.~Kundu, and R.~H.~Becker, Astron. Astrophys. \textbf{51}, 21 (1976).

\bibitem{38:Lozinskaya_n}
R.~Weaver, R.~McCray, J.~Castor, \emph{et al.}, Astrophys. J. \textbf{218}, 377 (1977).

\bibitem{37:Lozinskaya_n}
M.~J.~Whitehead, J.~Meaburn, and C.~A.~Claiton, Mon. Not. R. Astron. Soc. \textbf{237},
1109 (1989).

\bibitem{39:Lozinskaya_n}
F.P.~Wilkin, Astrophys. J. \textbf{459}, L3 (1996).

\end{thebibliography}
\end{document}